\documentclass[conference]{IEEEtran}

\usepackage{cite}
\usepackage{amsmath,amssymb,amsfonts}
\usepackage{graphicx}
\usepackage{textcomp}
\usepackage{xcolor}
\usepackage{url}
\usepackage{tikz}
\usetikzlibrary{
    arrows.meta,
    positioning,
    shapes.geometric,
    shapes.symbols,
    calc,
    fit,
    backgrounds
}

\definecolor{NormalBlue}{RGB}{35,86,140}
\definecolor{NormalFill}{RGB}{232,241,250}
\definecolor{AttackRed}{RGB}{176,48,48}
\definecolor{AttackFill}{RGB}{252,234,234}
\definecolor{DefenseGreen}{RGB}{42,124,72}
\definecolor{DefenseFill}{RGB}{233,247,237}
\definecolor{DecisionGold}{RGB}{158,112,20}
\definecolor{DecisionFill}{RGB}{253,247,225}
\definecolor{ImpactPurple}{RGB}{103,65,145}
\definecolor{ImpactFill}{RGB}{243,237,250}
\definecolor{NoteGray}{RGB}{90,90,90}
\definecolor{NoteFill}{RGB}{247,247,247}

\tikzset{
    normalblock/.style={
        draw=NormalBlue,
        fill=NormalFill,
        rounded corners=2pt,
        line width=0.8pt,
        align=center,
        text width=2.25cm,
        minimum height=1.05cm,
        font=\scriptsize
    },
    defenseblock/.style={
        draw=DefenseGreen,
        fill=DefenseFill,
        rounded corners=2pt,
        line width=0.9pt,
        align=center,
        text width=2.35cm,
        minimum height=1.12cm,
        font=\scriptsize
    },
    attackblock/.style={
        signal,
        signal to=east and west,
        draw=AttackRed,
        fill=AttackFill,
        line width=1pt,
        align=center,
        text width=2.2cm,
        minimum height=1.05cm,
        font=\scriptsize\bfseries
    },
    decisionblock/.style={
        diamond,
        draw=DecisionGold,
        fill=DecisionFill,
        line width=0.9pt,
        aspect=2.0,
        align=center,
        text width=1.45cm,
        inner sep=1pt,
        font=\scriptsize
    },
    impactblock/.style={
        draw=ImpactPurple,
        fill=ImpactFill,
        rounded corners=2pt,
        line width=0.8pt,
        align=center,
        text width=2.75cm,
        minimum height=1.12cm,
        font=\scriptsize
    },
    noteblock/.style={
        draw=NoteGray,
        fill=NoteFill,
        rounded corners=2pt,
        line width=0.6pt,
        align=left,
        text width=14.8cm,
        inner sep=4pt,
        font=\scriptsize
    },
    flowarrow/.style={
        -{Latex[length=2.1mm,width=1.5mm]},
        line width=0.8pt,
        draw=black
    },
    attackarrow/.style={
        -{Latex[length=2.1mm,width=1.5mm]},
        line width=1pt,
        dashed,
        draw=AttackRed
    },
    recoveryarrow/.style={
        -{Latex[length=2.1mm,width=1.5mm]},
        line width=0.9pt,
        dashed,
        draw=DefenseGreen
    },
    smalllabel/.style={
        font=\scriptsize,
        fill=white,
        inner sep=1pt,
        align=center
    }
}

\def\BibTeX{{\rm B\kern-.05em{\sc i\kern-.025em b}\kern-.08em
    T\kern-.1667em\lower.7ex\hbox{E}\kern-.125emX}}

\begin{document}
\bstctlcite{BSTcontrol}

\title{
Quantity-Level FDIA in SDR-Based P2P Energy Trading: Cyber-Economic Impact and Pre-Clearing Detection
}

\author{
\IEEEauthorblockN{Wassim Sarrieddine\IEEEauthorrefmark{1}, Razane Tajeddine\IEEEauthorrefmark{2}
}
\IEEEauthorblockA{
\IEEEauthorrefmark{1}\textit{Department of Electrical and Computer Engineering},
\textit{American University of Beirut},
Beirut, Lebanon
}
\IEEEauthorblockA{
\IEEEauthorrefmark{2}\textit{Department of Mathematics and Systems Analysis},
\textit{Aalto University},
Finland
}
emails: wzs01@mail.aub.edu, rstajeddine@gmail.com
}

\maketitle

\begingroup
\renewcommand{\thefootnote}{}
\footnotetext{\footnotesize
\copyright~2026 IEEE. Personal use of this material is permitted.
Permission from IEEE must be obtained for all other uses, in any
current or future media, including reprinting/republishing this
material for advertising or promotional purposes, creating new
collective works, for resale or redistribution to servers or lists,
or reuse of any copyrighted component of this work in other works.
}
\endgroup

\begin{abstract}
This paper investigates the cyber-economic impact and pre-clearing
detectability of false data injection attacks (FDIAs) in peer-to-peer
(P2P) energy trading among interconnected microgrids. In the proposed
cyber-economic framework, load and photovoltaic profiles drive a
sequential 15-minute market with deterministic double-sided proportional (pro-rata) allocation and supply-to-demand-ratio (SDR) pricing, while
true physical injections are separated from communicated market
quantities. The framework is evaluated through a multi-microgrid case
study over 69,904 market-aligned states, for which all AC power-flow
solutions converged and voltages remained within 0.98999--1.04369 per
unit. Healthy local exchange reached 1,159.66~MWh and generated
\$92,772.62 of community benefit. Bias, scaling, and ramp attacks modify
only one microgrid's communicated quantity, altering the settlement
price, P2P allocation, scheduled residual-grid exchange, participant
financial outcomes, and benefit fairness, while true feeder injections
remain unchanged. Role-specific random forest, statistical threshold,
and score-fusion detectors are compared using only information available
before clearing. Pooled sample-level detection rates range from 7.82\%
to 12.14\%, with false-alarm rates from 4.69\% to 7.70\% and role
dependence. In the investigated case study, measurable downstream economic
distortion and low pre-clearing detectability coexist under bounded
quantity manipulation.
\end{abstract}

\section{Introduction}
\label{sec:introduction}

The increasing penetration of distributed energy resources (DERs),
particularly photovoltaic (PV) generation, is transforming passive
consumers into active prosumers and creating new opportunities for
distribution-level energy exchange
\cite{tanis2025review,huang2026,liu2025p2p2g}. Because local renewable
generation and demand are rarely synchronized, one participant may
have excess PV generation while another simultaneously requires grid
energy. Peer-to-peer (P2P) trading addresses this mismatch by
enabling local exchange before residual energy is settled with the
utility grid, potentially improving renewable-energy utilization and
participant economics \cite{huang2026,zou2026}. Such operation,
however, increasingly depends on the integrity of the quantities and
prices communicated among participants and the market operator.

Recent P2P research has advanced auction-based allocation
\cite{huang2026,sim2024doubleauction}, privacy-preserving trading
\cite{zou2026}, dynamic supply--demand pricing
\cite{ghodusinejad2025}, and network-aware P2P-to-grid coordination
\cite{liu2025p2p2g}. While these approaches improve market efficiency
and coordination, they also increase dependence on communicated data;
a falsified quantity can therefore alter market outcomes even when the
participant's true load, generation, and feeder injections remain
unchanged. Cybersecurity studies have consequently examined false data injection
attacks (FDIAs) in local energy markets. Haggi and Sun \cite{haggi2022p2p} showed that malicious price
and quantity reporting can distort participant economics and, when
corrupted market decisions are physically implemented, feeder
operation. Mohammadi \emph{et al.} \cite{mohammadi2022fdia}
investigated machine-learning detection of P2P-specific FDIAs, while
subsequent work addressed compromised-data identification, resilient
local-energy operation, and learning-based anomaly detection
\cite{kermani2024xgboost,tabassum2024secured,
liu2025byzantine,luo2026}. More recently, Mofreh \emph{et al.}
\cite{mofreh2026} considered FDIA detection before market clearing
using information available at bid submission, highlighting the
difficulty of identifying subtle manipulation before downstream
effects become observable. 

A remaining gap is the joint assessment of \emph{attack consequence}
and \emph{pre-clearing observability}. Manipulating a communicated
market quantity leaves no physical signature in feeder variables, while
a pre-clearing detector cannot use the supply-to-demand ratio (SDR),
settlement price, allocation, participant cost, or other outcomes
generated after clearing. 

This paper
addresses this gap through a cyber-economic framework for P2P
energy trading among interconnected microgrids (MGs). The framework
separates true electrical quantities from communicated market information
and traces a targeted quantity-level FDIA through SDR formation, pricing,
allocation, residual-grid scheduling, participant economics, and benefit
distribution. The same attack is evaluated before clearing using only
the information available at that stage, so downstream economic impact
and pre-clearing observability are assessed on identical attacks. The framework is formulated independently of any specific feeder
topology or MG count and is evaluated through a distribution-network case
study.

The main contributions are:
\begin{itemize}
    \item A cyber-economic P2P trading framework that separates true
    electrical quantities from communicated market information, enabling
    attack-induced economic effects to be isolated from physical feeder
    disturbances.

    \item A quantity-level FDIA analysis that traces
    manipulation through SDR, pricing, allocation, residual-grid
    scheduling, participant economics, and benefit distribution across
    attack type, direction, severity, and market role.

    \item A leakage-safe pre-clearing detectability study comparing
    machine-learning, statistical, and score-fusion detectors under
    chronological held-out evaluation.
\end{itemize}

\section{Methodology}
\label{sec:methodology}

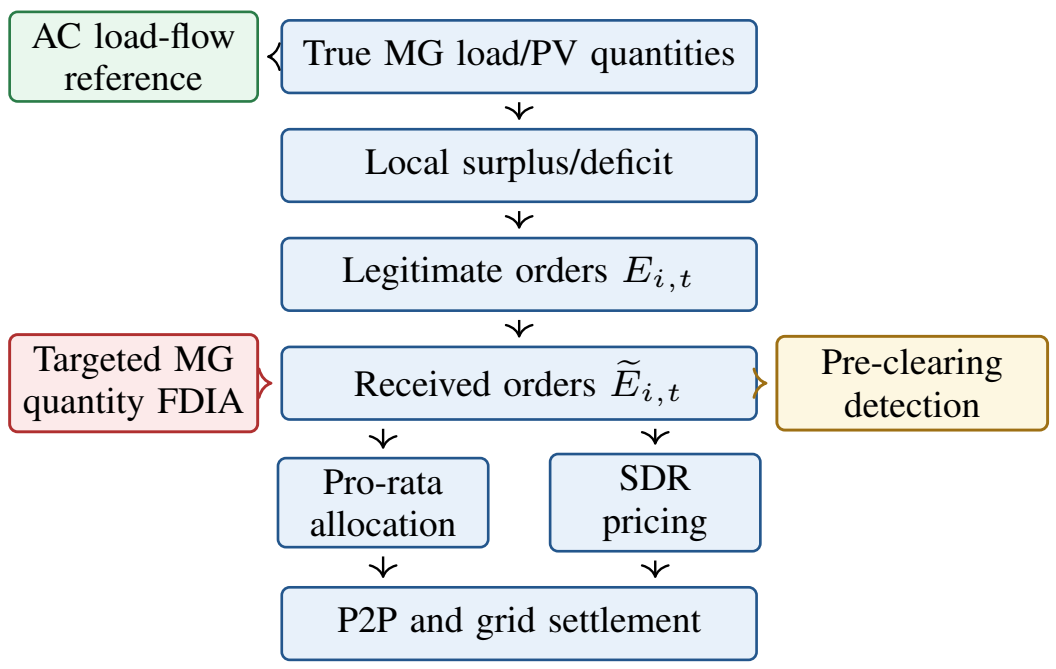
\begin{figure}[!b]
\centering

\begin{tikzpicture}[
    mainblock/.style={
        rectangle,
        draw=NormalBlue,
        fill=NormalFill,
        rounded corners=1.5pt,
        line width=0.5pt,
        text width=3.0cm,
        minimum height=0.48cm,
        align=center,
        inner sep=2pt,
        font=\scriptsize
    },
    physicalblock/.style={
        rectangle,
        draw=DefenseGreen,
        fill=DefenseFill,
        rounded corners=1.5pt,
        line width=0.5pt,
        text width=1.50cm,
        minimum height=0.50cm,
        align=center,
        inner sep=2pt,
        font=\scriptsize
    },
    attackblock2/.style={
        rectangle,
        draw=AttackRed,
        fill=AttackFill,
        rounded corners=1.5pt,
        line width=0.55pt,
        text width=1.50cm,
        minimum height=0.50cm,
        align=center,
        inner sep=2pt,
        font=\scriptsize
    },
    detectionblock/.style={
        rectangle,
        draw=DecisionGold,
        fill=DecisionFill,
        rounded corners=1.5pt,
        line width=0.55pt,
        text width=1.65cm,
        minimum height=0.62cm,
        align=center,
        inner sep=2pt,
        font=\scriptsize
    },
    smallblock/.style={
        rectangle,
        draw=NormalBlue,
        fill=NormalFill,
        rounded corners=1.5pt,
        line width=0.5pt,
        text width=1.25cm,
        minimum height=0.52cm,
        align=center,
        inner sep=2pt,
        font=\scriptsize
    }
]


\node[mainblock] (true) at (0,0)
{True MG load/PV quantities};

\node[mainblock] (balance) at (0,-0.72)
{Local surplus/deficit};

\node[mainblock] (order) at (0,-1.44)
{Legitimate orders $E_{i,t}$};

\node[mainblock] (received) at (0,-2.16)
{Received orders $\widetilde{E}_{i,t}$};

\draw[->, line width=0.45pt,
      shorten <=1.5pt, shorten >=1.5pt]
(true.south) -- (balance.north);

\draw[->, line width=0.45pt,
      shorten <=1.5pt, shorten >=1.5pt]
(balance.south) -- (order.north);

\draw[->, line width=0.45pt,
      shorten <=1.5pt, shorten >=1.5pt]
(order.south) -- (received.north);


\node[physicalblock] (ac) at (-2.55,0)
{AC load-flow\\reference};

\draw[->, line width=0.45pt,
      shorten <=1.5pt, shorten >=1.5pt]
(true.west) -- (ac.east);


\node[attackblock2] (attack) at (-2.55,-2.16)
{Targeted MG\\quantity FDIA};

\draw[->, dashed, draw=AttackRed,
      line width=0.55pt,
      shorten <=1.5pt, shorten >=1.5pt]
(attack.east) -- (received.west);

\node[detectionblock] (detect) at (2.60,-2.16)
{Pre-clearing\\detection};

\draw[->, dashed, draw=DecisionGold,
      line width=0.55pt,
      shorten <=1.5pt, shorten >=1.5pt]
(received.east) -- (detect.west);


\node[smallblock] (auction) at (-0.90,-2.95)
{Pro-rata\\allocation};

\node[smallblock] (sdr) at (0.90,-2.95)
{SDR\\pricing};

\draw[->, line width=0.45pt,
      shorten <=1.5pt, shorten >=1.5pt]
([xshift=-0.90cm]received.south) -- (auction.north);

\draw[->, line width=0.45pt,
      shorten <=1.5pt, shorten >=1.5pt]
([xshift=0.90cm]received.south) -- (sdr.north);


\node[mainblock] (settlement) at (0,-3.75)
{P2P and grid settlement};

\draw[->, line width=0.45pt,
      shorten <=1.5pt, shorten >=1.5pt]
(auction.south) -- ([xshift=-0.90cm]settlement.north);

\draw[->, line width=0.45pt,
      shorten <=1.5pt, shorten >=1.5pt]
(sdr.south) -- ([xshift=0.90cm]settlement.north);

\end{tikzpicture}

\caption{Cyber-economic P2P workflow with FDIA and pre-clearing detection.}
\label{fig:methodology_flow}
\end{figure}

Legitimate orders are formed from local demand and generation, after
which an FDIA may alter a targeted communicated quantity, and detection
is performed before clearing. Because
healthy and attacked cases retain the same electrical state, market and
economic deviations isolate the effect of information manipulation.
The workflow is summarized in Fig.~\ref{fig:methodology_flow}.


\subsection{Electrical Reference}
\label{subsec:system}

Consider a distribution network with participating MGs
$\mathcal{M}=\{1,\ldots,N\}$ and electrical nodes $\mathcal{N}$.
For MG $i$ at time $t$, $P_{i,t}^{L}$ and $P_{i,t}^{G}$ denote true
active load and generation, with corresponding reactive quantities
$Q_{i,t}^{L}$ and $Q_{i,t}^{G}$. These quantities determine the
electrical state independently of the information communicated to the market and satisfy the standard alternating-current (AC) nodal
power-balance relation 
\cite{liu2025p2p2g},
\begin{equation}
\begin{aligned}
S_{n,t}^{\mathrm{inj}}
&=
\left(P_{n,t}^{G}-P_{n,t}^{L}\right)
+\mathrm{j}\left(Q_{n,t}^{G}-Q_{n,t}^{L}\right)\\
&=
V_{n,t}
\left(
\sum_{m\in\mathcal{N}}Y_{nm}V_{m,t}
\right)^{*},
\end{aligned}
\label{eq:ac_reference}
\end{equation}
where $P_{n,t}^{G}$, $P_{n,t}^{L}$, $Q_{n,t}^{G}$, and
$Q_{n,t}^{L}$ are the aggregate generation and load quantities connected
at node $n$, $S_{n,t}^{\mathrm{inj}}$ is the resulting complex net
injection, $V_{n,t}$ is the bus voltage, $Y_{nm}$ is the
admittance-matrix element, and $(\cdot)^*$ denotes complex conjugation. The AC solution provides the
electrical reference for voltage, losses, grid exchange, reverse flow,
and equipment utilization. Under the threat model, falsified market data do not alter the true
injections, and the AC solution neither constrains nor re-clears the
P2P market; attack effects are therefore confined to the cyber-economic
layer.

\subsection{P2P Trading and SDR-Based Pricing}
\label{subsec:trading}

For each interval $\Delta t$, local generation first serves local
demand. Self-consumed power and residual surplus and deficit energy are
\cite{huang2026,sim2024doubleauction,zou2026}
\begin{equation}
\begin{aligned}
P_{i,t}^{\mathrm{SC}}
&=\min\!\left(P_{i,t}^{G},P_{i,t}^{L}\right),\\
E_{i,t}^{\mathrm{sur}}
&=\max\!\left(P_{i,t}^{G}-P_{i,t}^{L},0\right)\Delta t,\\
E_{i,t}^{\mathrm{def}}
&=\max\!\left(P_{i,t}^{L}-P_{i,t}^{G},0\right)\Delta t.
\end{aligned}
\label{eq:mg_balance}
\end{equation}

{Using a tolerance $\epsilon$, MG $i$ joins the seller set
$\mathcal{S}_t$ if $E_{i,t}^{\mathrm{sur}}>\epsilon$ and the buyer set
$\mathcal{B}_t$ if $E_{i,t}^{\mathrm{def}}>\epsilon$; otherwise, it is
balanced and does not trade. The legitimate market order is
$E_{i,t}=E_{i,t}^{\mathrm{sur}}$ for $i\in\mathcal{S}_t$ and
$E_{i,t}=E_{i,t}^{\mathrm{def}}$ for $i\in\mathcal{B}_t$, while
$\widetilde E_{i,t}$ denotes the corresponding order received by the
market operator, with $\widetilde E_{i,t}=E_{i,t}$ under healthy
communication. Seller asks and buyer bids equal the grid-export tariff
$\lambda_s^G$ and grid-import tariff $\lambda_b^G$, respectively.

Aggregate communicated supply, demand, and SDR are
\begin{equation}
\begin{aligned}
S_t &=
\sum_{i\in\mathcal{S}_t}\widetilde E_{i,t},
\qquad
D_t =
\sum_{i\in\mathcal{B}_t}\widetilde E_{i,t},\\
\mathrm{SDR}_t &= \frac{S_t}{D_t},
\qquad S_t,D_t>0.
\end{aligned}
\label{eq:sdr}
\end{equation}

If either $S_t$ or $D_t$ is zero, no P2P exchange is cleared, no
internal P2P settlement price is required, and the residual energy is
scheduled with the utility grid. Otherwise,
$E_t^{\mathrm{P2P}}=\min(S_t,D_t)$ is distributed among equal-priority
participants by deterministic double-sided proportional (pro-rata) allocation \cite{sim2024doubleauction,huang2026}:
\begin{equation}
E_{i,t}^{\mathrm{clr}}=
\begin{cases}
\dfrac{\widetilde E_{i,t}}{S_t}E_t^{\mathrm{P2P}},
& i\in\mathcal{S}_t,\\[6pt]
\dfrac{\widetilde E_{i,t}}{D_t}E_t^{\mathrm{P2P}},
& i\in\mathcal{B}_t.
\end{cases}
\label{eq:proportional_allocation}
\end{equation}
For $S_t,D_t>0$, the settlement price follows the supply--demand
condition using SDR-based pricing \cite{ghodusinejad2025,zeyad2026}:
\begin{equation}
\lambda_t=
\lambda_{\mathrm{bal}}
+
\lambda_{\mathrm{ctrl}}
\tanh\!\left[-t_f\ln(\mathrm{SDR}_t)\right],
\label{eq:sdr_price}
\end{equation}
where
$\lambda_{\mathrm{bal}}=(\lambda_b^G+\lambda_s^G)/2$,
$\lambda_{\mathrm{ctrl}}=(\lambda_b^G-\lambda_s^G)/2$, and $t_f>0$
controls price sensitivity; $\mathrm{SDR}_t<1$ moves the price toward
the grid-import tariff, whereas $\mathrm{SDR}_t>1$ moves it toward the
grid-export tariff. Since $\lambda_s^G<\lambda_t<\lambda_b^G$, every
cleared unit is individually beneficial to both sides relative to
direct grid settlement; hence, the full coincident quantity
$E_t^{\mathrm{P2P}}=\min(S_t,D_t)$ is matched independently of the
price. Allocation (\ref{eq:proportional_allocation}) determines the
traded energy, whereas $\lambda_t$ determines how the resulting surplus
is shared among participants.

Unmatched communicated quantities are settled with the utility grid,
giving aggregate residual import
$E_t^{\mathrm{G,imp}}=D_t-E_t^{\mathrm{P2P}}$ and export
$E_t^{\mathrm{G,exp}}=S_t-E_t^{\mathrm{P2P}}$, with
$E_t^{\mathrm{P2P}}=0$ when no exchange is cleared. With no transaction
fee, the net settlement cost of MG $i$ (negative values denote net
revenue) for a cleared P2P interval is
\begin{equation}
C_{i,t}=
\begin{cases}
-\lambda_t E_{i,t}^{\mathrm{clr}}
-\lambda_s^G\!\left(\widetilde E_{i,t}-E_{i,t}^{\mathrm{clr}}\right),
& i\in\mathcal{S}_t,\\[4pt]
\lambda_t E_{i,t}^{\mathrm{clr}}
+\lambda_b^G\!\left(\widetilde E_{i,t}-E_{i,t}^{\mathrm{clr}}\right),
& i\in\mathcal{B}_t.
\end{cases}
\label{eq:settlement_cost}
\end{equation}
When no P2P exchange is cleared,
$C_{i,t}=-\lambda_s^G\widetilde E_{i,t}$ for sellers and
$C_{i,t}=\lambda_b^G\widetilde E_{i,t}$ for buyers; balanced MGs have
$C_{i,t}=0$. When P2P exchange is cleared, the benefit relative to direct grid
settlement is
$B_{i,t}=E_{i,t}^{\mathrm{clr}}(\lambda_t-\lambda_s^G)$ for sellers and
$B_{i,t}=E_{i,t}^{\mathrm{clr}}(\lambda_b^G-\lambda_t)$ for buyers;
otherwise, $B_{i,t}=0$.
Summing these benefits gives
\begin{equation}
B_t^{\mathrm{com}}
=
E_t^{\mathrm{P2P}}
\left(\lambda_b^G-\lambda_s^G\right),
\label{eq:community_benefit}
\end{equation}
so, for fixed matched energy, the SDR price redistributes surplus among
participants but does not change the total community benefit. Its distribution across the
$N$ MGs is quantified by Jain's fairness index:
\begin{equation}
J=
\frac{\left(\sum_{i=1}^{N}B_i\right)^2}
{N\sum_{i=1}^{N}B_i^2},
\qquad
B_i=\sum_t B_{i,t}.
\label{eq:jain}
\end{equation}

\subsection{Cyberattack and Pre-Clearing Detection}
\label{subsec:cyber}

\subsubsection{Quantity-Level FDIA}

The threat model considers a third-party adversary on the communication
path between a targeted MG and the market operator. After legitimate
order formation, only the communicated market-energy magnitude is
modified; true load, generation, reactive demand, timestamp, market
role, and ask/bid price remain unchanged
\cite{haggi2022p2p,mohammadi2022fdia}. A falsified seller quantity
perturbs $S_t$, whereas a falsified buyer quantity perturbs $D_t$,
thereby affecting the SDR, settlement price, allocation, residual-grid
scheduling, and participant economics without changing the electrical
state. Let $j\in\mathcal{M}$ denote the targeted MG,
$r\in\{\mathrm{S},\mathrm{B}\}$ its market role, and $E_{\mathrm{ref},r}$ an attack-free
role-specific reference quantity. For an attack episode beginning at
$t_0$, let $t=t_0+\ell-1$, with $\ell=1,\ldots,L$. The bias, scaling,
and ramp attacks are defined as
\begin{equation}
\widetilde E_{j,t}=
\begin{cases}
E_{j,t}+d\,sE_{\mathrm{ref},r},
& \mathrm{Bias},\\[3pt]
(1+d\,s)E_{j,t},
& \mathrm{Scaling},\\[3pt]
E_{j,t}+\dfrac{\ell}{L}d\,sE_{\mathrm{ref},r},
& \mathrm{Ramp},
\end{cases}
\label{eq:fdia}
\end{equation}
where $d\in\{+1,-1\}$ denotes over- or under-reporting and $s>0$ the
attack severity. All non-targeted orders remain unchanged,
$\widetilde E_{i,t}=E_{i,t}$ for $i\neq j$. Active communicated orders must be finite and positive. For the tested
$s\leq0.2$, scaling always satisfies this, whereas bias and ramp
under-reporting require $E_{j,t}>sE_{\mathrm{ref},r}$ in the worst
case. Evaluated windows are pre-screened accordingly; non-positive
orders would be rejected rather than clipped, and all 216 attacked MG2
orders per scenario pass validation.

\subsubsection{Pre-Clearing Detection}

Detection is performed when the communicated order reaches the market
operator, before clearing. Only current and past information is used;
future observations, post-clearing variables (SDR, price, allocation,
and participant cost), and AC-network variables are excluded to prevent
downstream information leakage
\cite{mohammadi2022fdia,mofreh2026}. For role $r$, the feature vector
$\mathbf{x}_t^{(r)}$ is constructed from the received order
$\widetilde E_{j,t}$ and prior same-role observations, capturing
temporal behavior, expected-order residuals, seasonal patterns, and
cumulative deviations. A role-specific random forest (RF) captures
nonlinear feature interactions without a parametric model and provides
probability scores for fusion
\cite{mohammadi2022fdia,kermani2024xgboost}. Chronological
training--validation--test separation prevents held-out information
from entering calibration. The RF decision is
\begin{equation}
A_t^{\mathrm{RF}}
=
\mathbb{I}\!\left[
p_t^{(r)}\geq\tau_{\mathrm{RF},r}
\right],
\label{eq:rf_detector}
\end{equation}
where $\mathbb{I}[\cdot]$ denotes the indicator function,
$p_t^{(r)}$ is the estimated attack probability,
$\tau_{\mathrm{RF},r}$ is the validation-selected threshold, and
$A_t^{\mathrm{RF}}=1$ denotes an alarm.
A complementary statistical detector uses the seasonal robust score
\begin{equation}
z_{t,r}^{\mathrm{sea}}
=
\frac{\widetilde E_{j,t}-\widehat E_{t,r}^{\mathrm{sea}}}
{1.4826\,\mathrm{MAD}_{t,r}^{\mathrm{sea}}},
\label{eq:seasonal_score}
\end{equation}
where $\widehat E_{t,r}^{\mathrm{sea}}$ and
$\mathrm{MAD}_{t,r}^{\mathrm{sea}}$ are the attack-free seasonal median
and median absolute deviation (MAD); 1.4826 provides normal-consistency
scaling \cite{paudel2024}. Robust-residual and cumulative-sum components
were also evaluated on validation data but not retained. Thus,
$h_t^{(r)}=|z_{t,r}^{\mathrm{sea}}|/\tau_{\mathrm{sea},r}$ and the
statistical alarm is
\begin{equation}
A_t^{\mathrm{TH}}
=
\mathbb{I}\!\left[h_t^{(r)}>1\right].
\label{eq:threshold_alarm}
\end{equation}

Score-level fusion normalizes each branch by its own alarm threshold,
so unity marks that branch's alarm point, and combines the resulting
evidence without hard gating:
\begin{equation}
\begin{aligned}
\rho_t^{(r)}
&=\frac{p_t^{(r)}}{\tau_{\mathrm{RF},r}},\\
c_t^{(r)}
&=\alpha_r\rho_t^{(r)}
+(1-\alpha_r)h_t^{(r)},\\
A_t^{\mathrm{FUS}}
&=\mathbb{I}\!\left[
c_t^{(r)}\geq\tau_{\mathrm{FUS},r}
\right].
\end{aligned}
\label{eq:Fusion_detector}
\end{equation}
Here, $\alpha_r\in[0,1]$ is the role-specific fusion weight and
$\tau_{\mathrm{FUS},r}$ the fused-score threshold. Alarms are diagnostic
only and do not alter market clearing.
\section{Case Study and Simulation Results}
\label{sec:results}

\subsection{Case-Study Configuration}
\label{subsec:case_configuration}

The framework of Section~\ref{sec:methodology} is evaluated using four
MGs on a reduced balanced positive-sequence IEEE 13-node feeder
\cite{kersting2001}, retaining the source, transformer, regulator, and
MG buses. MG1--MG4 are connected at nodes 632, 671, 680, and 675,
respectively; MG1--MG3 include load and PV, while MG4 is load-only.
The 4.16-kV feeder is supplied from node~650. This case-specific
configuration does not restrict the general framework. Minute-resolution load
and solar profiles are obtained from PSML \cite{zheng2022psml}, with
PJM zone~1, CAISO zone~4, ERCOT zone~3, and PJM zone~20 assigned to
MG1--MG4, respectively, to provide diverse temporal behavior without
implying geographical co-location. Normalized load and solar profiles, $\varphi_i^{L}(t)$ and
$\varphi_i^{G}(t)$, are converted to physical power as
\begin{equation}
\begin{aligned}
P_{i,t}^{L} &= \kappa W_i\varphi_i^{L}(t), \qquad
P_{i,t}^{G}=P_i^{\mathrm{PV,rated}}\varphi_i^{G}(t),\\
\kappa &=
\frac{P_{\mathrm{ref}}}
{\displaystyle\max_t\sum_i W_i\varphi_i^{L}(t)},
\end{aligned}
\label{eq:case_scaling}
\end{equation}
where $W=[1.000,1.155,0.298,1.013]$~MW,
$P^{\mathrm{PV,rated}}=[1.000,1.155,0.298,0]$~MW, and
$P_{\mathrm{ref}}=3.466$~MW. PV is not clipped to local demand,
allowing surplus operation. Loads operate at 0.95 lagging power factor,
PV at unity power factor, and no supplementary reactive-power support
is used. Newton--Raphson AC power flow is solved with the true injections for
all 69,904 15-min states of the two-year dataset, with voltage assessed
against 0.95--1.05~p.u.
The 15-min P2P market uses realized quantities, a role tolerance
$\epsilon=10^{-4}$~MWh, grid-export and grid-import tariffs of
\$40/MWh and \$120/MWh, respectively, and $t_f=1$
(sensitivity: $t_f\in\{0.5,1,2\}$), yielding
$\lambda_{\mathrm{bal}}=80$~\$/MWh and
$\lambda_{\mathrm{ctrl}}=40$~\$/MWh.

MG2 is targeted in both seller and buyer operation. Bias, scaling, and
ramp attacks are applied in over- and under-reporting directions at
5\%, 10\%, and 20\% severity, yielding 18 scenarios. Each episode
contains $L=12$ consecutive 15-min samples. Bias and ramp use $E_{\mathrm{ref},\mathrm{S}}=0.050253$~MWh and $E_{\mathrm{ref},\mathrm{B}}=0.128848$~MWh, while scaling acts directly on the
instantaneous quantity. The held-out schedule contains eighteen
non-overlapping 3-h episodes, with nine seller and nine buyer windows,
identical across scenarios and totaling 54~h of attacked operation per
scenario. The 69,904 intervals are divided chronologically into 41,942 training,
13,981 validation, and 13,981 held-out intervals; training attacks are
generated independently of the held-out schedule. Features use a
96-step same-role history, four expanding chronological folds, and a
48-step embargo. Training-only ablation retains 71 raw order-derived
features, yielding 127 RF inputs after imputation and missingness
indicators. Seller and buyer RFs use 400 trees with maximum depths of
8 and 16, respectively. Validation selects
$\tau_{\mathrm{RF},\mathrm{S}}=0.4429$,
$\tau_{\mathrm{RF},\mathrm{B}}=0.2601$,
$\tau_{\mathrm{sea},\mathrm{S}}=2.4672$, and
$\tau_{\mathrm{sea},\mathrm{B}}=3.4759$. The fusion search over $\alpha_r\in\{0.1,0.2,\ldots,0.9\}$ selects
$\alpha_{\mathrm{S}}=0.8$, $\alpha_{\mathrm{B}}=0.6$,
$\tau_{\mathrm{FUS},\mathrm{S}}=1.1421$, and
$\tau_{\mathrm{FUS},\mathrm{B}}=1.0579$.
Role-specific RF/statistical thresholds and each fusion pair
$(\alpha_r,\tau_{\mathrm{FUS},r})$ maximize the mean validation
detection rate (DR) across scenarios subject to a mean validation
false-alarm rate (FAR) $\mathrm{FAR}_{\mathrm{val}}\leq5\%$, with
mean F1 as a tie-breaker, and are then frozen for held-out evaluation.

For reproducibility, the implementation notebook and environment
requirements are publicly available in the project repository.%
\footnote{\url{https://github.com/wassim-sarrieddine/secure-p2p-energy-trading}}

\subsection{Electrical Reference and Healthy P2P Market}
\label{subsec:results_healthy}

The 69,904 market-aligned AC states establish the electrical reference.
All solutions converged without voltage violations; voltage,
point-of-common-coupling (PCC), regulator, and loss results are
summarized in Table~\ref{tab:healthy_results}.
Feeder-wide thermal compliance is not claimed where verified ampacity
data are unavailable. Healthy P2P operation exchanged 1,159.66~MWh over 19,477 active
intervals and generated \$92,772.62 of community benefit relative to
direct grid settlement. Among active intervals, 86.33\% satisfy $\mathrm{SDR}_t<1$, giving a
P2P-energy-weighted mean price of \$94.71/MWh, above the \$80/MWh
balancing reference and toward the grid-import tariff
\cite{ghodusinejad2025}. Nearly all time-coincident matchable energy is
exchanged locally, although P2P occurs during only 27.86\% of the
horizon. The Jain index of 0.8352 indicates a relatively balanced but
nonuniform benefit distribution. Varying $t_f$ from 0.5 to 2 changes prices and benefit distribution,
not matched energy or community benefit. 
\begin{table}[!b]
\caption{Healthy electrical and P2P market results}
\label{tab:healthy_results}
\centering
\small
\renewcommand{\arraystretch}{0.95}
\setlength{\tabcolsep}{2.6pt}
\begin{tabular}{l r}
\hline
\textbf{Metric} & \textbf{Result} \\
\hline
Converged 15-min AC states & 69,904/69,904 \\
Voltage range & 0.98999--1.04369 p.u. \\
Voltage violations & 0 \\
Maximum PCC import/export & 3.34/0.80 MW \\
Regulator tap range & 4--7 \\
Maximum regulator loading & 71.93\% \\
Active-energy losses & 279.13 MWh \\
P2P scheduled energy & 1,159.66 MWh \\
Active P2P intervals & 19,477 (27.86\%) \\
Residual grid purchase/export & 23,290.85/152.47 MWh \\
Community benefit & \$92,772.62 \\
P2P-energy-weighted mean price & \$94.71/MWh \\
Benefit-distribution Jain index & 0.8352 \\
\hline
\end{tabular}
\end{table}

\subsection{Cyber-Economic Impact of Quantity-Level FDIA}
\label{subsec:results_attack}

The held-out windows are re-evaluated with only MG2's communicated
quantity modified. All 18 scenarios use identical windows, while true
load, PV generation, market role, and feeder injections remain
unchanged. Define
$\Delta\lambda_t=\lambda_t^{A}-\lambda_t^{H}$ and
$\Delta\mathrm{SDR}_t=\mathrm{SDR}_t^{A}-\mathrm{SDR}_t^{H}$.
The term $|\Delta E^{\mathrm{P2P}}|$ measures the change in scheduled
matched energy,
while allocation and residual-grid distortions measure P2P and
utility-grid schedule changes. Participant financial distortion is
$\sum_i|C_i^{A}-C_i^{H}|$, where
$C_i=\sum_t C_{i,t}$ from (\ref{eq:settlement_cost}) is evaluated on
the scheduled communicated quantities. Community-benefit degradation is
$\sum_i(C_i^{A}-C_i^{H})$, equivalently the loss of aggregate benefit
because healthy and attacked benefits use the same grid-only reference.
Jain degradation uses the healthy same-window benefit distribution.
Table~\ref{tab:attack_results} summarizes the largest impacts. No single attack dominates all consequences. Scaling under-reporting
produces the largest SDR, price, and matched-energy deviations, whereas
bias over-reporting produces the largest residual-grid and participant
financial distortions. Bias under-reporting causes the strongest
fairness degradation: at 20\% severity, the same-window Jain index
falls from 0.7479 to 0.6102, so the full-horizon value of 0.8352 is not
the appropriate baseline for this window-specific metric.

Fig.~\ref{fig:cy1_propagation} illustrates the propagation mechanism.
During a representative 20\% scaling under-reporting seller episode,
MG2's communicated energy decreases from 0.05651 to 0.04521~MWh,
reducing SDR from 0.4195 to 0.3356 and increasing the settlement price
from \$108.03/MWh to \$111.90/MWh. The maximum episode price deviation
is \$6.547/MWh. Seller over-reporting produces the opposite price
direction, while buyer operation reverses the response. Mean absolute
price distortion also increases monotonically with severity for the
tested profile--direction combinations. Attack sensitivity depends strongly on market role. For 20\% scaling
under-reporting, mean absolute price deviation is approximately
\$2.82/MWh in seller episodes versus \$0.51/MWh in buyer episodes.
Because the roles occur under different operating conditions, this
difference reflects role-conditioned sensitivity rather than a causal
effect of role alone. Small falsified energy volumes can nevertheless
produce measurable settlement effects: 20\% bias over-reporting
accumulates 3.869~MWh of falsification and causes \$264.87 of
participant financial distortion (\$68.47/falsified MWh), while the
largest normalized value is \$69.30/falsified MWh for 20\% ramp
over-reporting. Some under-reporting scenarios increase the signed aggregate
community-benefit metric while still distorting price, allocation, grid
schedule, or fairness. A favorable aggregate metric therefore does not imply a secure or
equitable market.
\begin{table}[!t]
\caption{Largest held-out FDIA impacts over the 54-h attacked schedule}
\label{tab:attack_results}
\centering
\small
\renewcommand{\arraystretch}{0.95}
\setlength{\tabcolsep}{1.2pt}

\begin{tabular*}{\columnwidth}{@{\extracolsep{\fill}} l l r @{}}
\hline
\textbf{Impact metric} & \textbf{Scenario} & \textbf{Value} \\
\hline
Mean $|\Delta\lambda|$
 & Scaling under 20\% & \$1.662/MWh \\
Maximum $|\Delta\lambda|$
 & Scaling under 20\% & \$6.547/MWh \\
Mean $|\Delta\mathrm{SDR}|$
 & Scaling under 20\% & 0.0407 \\
$|\Delta E^{\mathrm{P2P}}|$, scheduled
 & Scaling under 20\% & 1.157 MWh \\
\hline
P2P allocation distortion
 & Bias under 20\% & 1.757 MWh \\
Residual-grid distortion
 & Bias over 20\% & 1.883 MWh \\
Participant financial distortion
 & Bias over 20\% & \$264.87 \\
Community-benefit degradation
 & Bias over 20\% & \$211.14 \\
Jain-index degradation
 & Bias under 20\% & 0.1377 \\
\hline
\end{tabular*}
\end{table}

\begin{figure*}[!t]
\centering
\includegraphics[width=0.82\textwidth]
{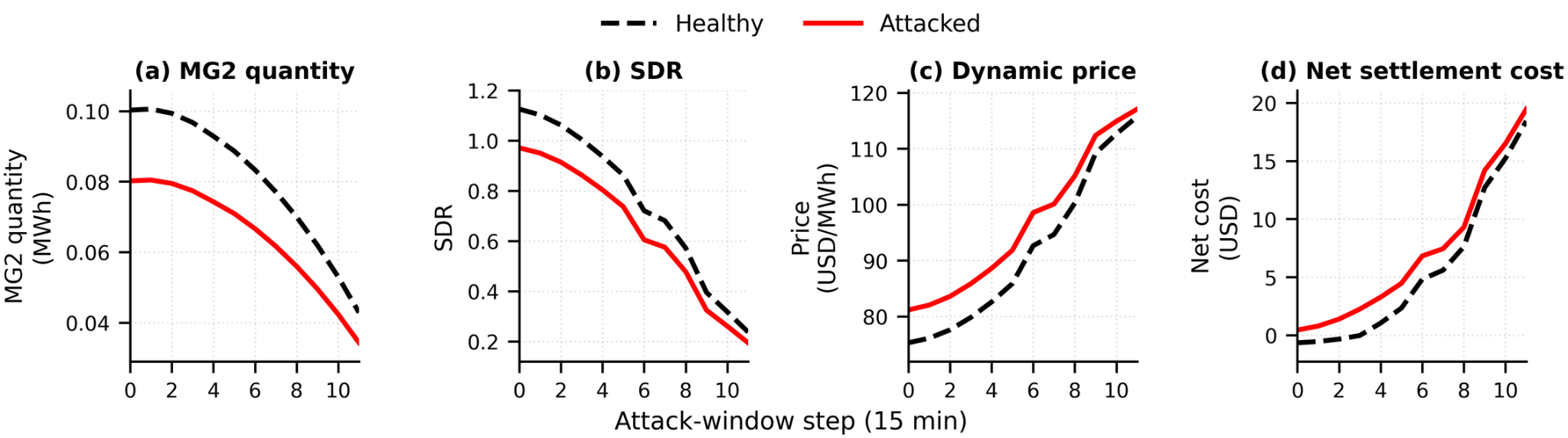}
\caption{Propagation of a 20\% scaling under-reporting FDIA in an MG2
seller episode across quantity, SDR, price, and community net
settlement cost.}
\label{fig:cy1_propagation}
\end{figure*}
For power-system operation, the 1.883~MWh residual-grid distortion is a
schedule error rather than a change in true feeder injections. Hence,
voltage or loading monitoring alone cannot verify market integrity.
If P2P schedules are translated into physical dispatch, corrupted
market information may propagate to feeder operation
\cite{haggi2022p2p}; that coupling is excluded here to isolate the
cyber-economic pathway.

\subsection{Pre-Clearing FDIA Detection}
\label{subsec:results_detection}
The same attacks are examined before clearing, when SDR, price,
allocation, and financial outcomes are unavailable. DR and FAR are the fractions of attacked and normal samples flagged,
respectively. Each scenario contains 108 attacked and
2,572 normal seller samples and 108 attacked and 11,192 normal buyer
samples; across 18 scenarios this gives 1,944 attacked evaluations
per role. Because attacks are sparse, the F1 score, the area under the receiver operating
characteristic curve (ROC-AUC; 0.5 is random), and average-precision
(AP) lift (AP divided by attack prevalence; no-skill
value 1) are reported instead of accuracy. Descriptive 95\% Wilson
half-widths are at most 1.9 percentage points for DR and 1.3 for FAR,
using unique normal-role samples; they do not account for temporal or
repeated-window dependence.

Pooled across roles, the fusion, RF, and threshold detectors achieve
DR/FAR values
of 7.82/4.69\%, 10.98/7.70\%, and 12.14/5.01\%, respectively. The
role-specific results in Table~\ref{tab:detection_results} are more
informative: RF is strongest for sellers (20.06\% DR), whereas the
threshold detector is strongest for buyers (23.25\% DR at 3.36\% FAR).
Conversely, the seller threshold detector performs below chance
(ROC-AUC 0.4326; AP lift 0.80), while buyer RF is near chance
(ROC-AUC 0.5361; AP lift 1.04). Although all settings satisfy the mean validation FAR constraint,
several held-out FARs exceed 5\%, reflecting chronological distribution
shift between validation and test periods. Static fusion does not preserve both role-specific strengths: for
sellers, it lowers FAR from 9.02\% (RF) to 2.70\% but reduces DR from
20.06\% to 8.85\%; for buyers, DR falls from 23.25\% (threshold) to
6.79\%, motivating role-gated or learned fusion.

The modest DR should be interpreted in light of the deliberately
restrictive pre-clearing information boundary. Each attack perturbs only
one communicated MG quantity by 5--20\% while preserving market role,
timestamp, ask/bid price, and true load and generation. Consequently,
the attack creates neither a role inconsistency nor an electrical
signature in the AC variables. Moreover, several downstream variables affected by the attack,
including SDR, settlement price, allocation, residual-grid schedule,
and participant cost, are generated only after clearing and are
intentionally excluded from detection. The detector must
therefore distinguish bounded falsification from normal order variability
using only current and historical same-role behavior.

\begin{table}[!b]
\caption{Role-specific held-out pre-clearing detection}
\label{tab:detection_results}
\centering
\small
\renewcommand{\arraystretch}{0.95}
\setlength{\tabcolsep}{1.45pt}
\begin{tabular}{lccccc}
\hline
\textbf{Method} &
\textbf{DR (\%)} &
\textbf{FAR (\%)} &
\textbf{F1 score} &
\textbf{ROC-AUC} &
\textbf{AP lift} \\
\hline
\multicolumn{6}{l}{\textbf{Seller}} \\
RF        & 20.06 & 9.02  & 0.1199 & 0.5742 & 2.38 \\
Threshold & 1.03  & 12.21 & 0.0053 & 0.4326 & 0.80 \\
Fusion    & 8.85  & 2.70  & 0.1023 & 0.5384 & 1.50 \\
\hline
\multicolumn{6}{l}{\textbf{Buyer}} \\
RF        & 1.90  & 7.40 & 0.0044 & 0.5361 & 1.04 \\
Threshold & 23.25 & 3.36 & 0.0986 & 0.6205 & 5.26 \\
Fusion    & 6.79  & 5.14 & 0.0212 & 0.6174 & 1.43 \\
\hline
\end{tabular}
\end{table}
These results are consistent with prior work, although attack models and
evaluation settings differ: Paudel~\cite{paudel2024} reports 4.65\%
recall for a MAD detector under small random-offset attacks, and Mofreh
\emph{et al.}~\cite{mofreh2026} identify subtle load/generation
manipulation as the hardest pre-clearing attack class.

In the investigated case study, measurable economic distortion and low
pre-clearing detectability coexist: 20\% scaling under-reporting
produced the largest reported price and SDR deviations, while pooled
sample-level DRs of the evaluated detectors were at most 12.14\%.
No general or causal relationship between distortion severity and
detectability is implied.

\subsection{Limitations and Scope}
\label{subsec:limitations}
The quantitative evaluation considers one feeder/market case, one
targeted MG, role-preserving single-variable manipulation, and three
bounded attack profiles. The AC network serves only as an electrical
reference; corrupted market schedules are not translated into physical
redispatch, so cyber-physical propagation is outside the present scope.
Independent capacity-based order caps and imbalance settlement are not
modeled; reported economic impacts are therefore schedule-based.
Detection is diagnostic and does not block, reconstruct, or mitigate
falsified orders. In addition, the restrictive pre-clearing information
boundary limits detector observability and the generalizability of its
numerical performance. Scenario-level consequence--detectability analysis, multi-point and
adaptive attacks, dispatch coupling, learned fusion, and mitigation and
recovery remain for future work.
\section{Conclusion}
\label{sec:conclusion}

This paper investigated the cyber-economic impact and pre-clearing
detectability of quantity-level FDIAs in SDR-based P2P trading among
interconnected microgrids. Separating true electrical quantities from
communicated market information isolated attack-induced economic
effects, while true physical injections remained unchanged. The
investigated case study showed distortions in SDR, price, allocation,
residual-grid scheduling, participant economics, and benefit fairness.
Detectability was role-dependent: RF performed best for sellers, the
statistical threshold for buyers, and static fusion did not preserve
both strengths. In this case study, measurable downstream economic
distortion and low pre-clearing detectability coexisted under bounded
single-variable manipulation.

\bibliographystyle{IEEEtran}
\bibliography{references}

\end{document}